# Removal of guided acoustic wave Brillouin scattering to the quantum-noise limit using symmetric interferometry in twisted photonic crystal fibers

Vishal Choudhury[1*c], Kevin Jaksch[1,2*], Markus Lippl[1,2], Gerd Leuchs[1,2],
Christoph Marquardt[2,1] and Nicolas Y. Joly[2,1†]
[1]Max Planck Institute for the Science of Light, Staudtstrasse 2, 91058 Erlangen, Germany
[2]Department of Physics, Friedrich-Alexander-Universität, 91058 Erlangen, Germany
†nicolas.joly@fau.de, [c]vishal.choudhury@mpl.mpg.de
*These authors contributed equally

Guided acoustic wave Brillouin scattering (GAWBS) is a major obstacle in fiber-based quantum and high-speed classical communication systems as well as in interferometry. The transverse phonons driving it modulate the light field in the fiber core, adding thermal noise to the signal. To this day, there is no known method to eliminate GAWBS from the fiber or to compensate for its effects completely. In this letter, we present twisted photonic crystal fibers (t-PCF) as the first-ever fiber system allowing a complete removal of mixed torsional radial GAWBS in a Stokes basis. The torsional radial modes modulate the fiber asymmetrically in the transverse direction, resulting in linear birefringence. While pure phase modulation is added as common noise in the guided fiber modes and can be easily removed through self-referencing, linear birefringence induces polarization modulation, which cannot be counteracted. In t-PCFs, the transverse symmetry of the geometry translates to multiple symmetries in the acoustic and optical domains in the circular basis. This enables equal phase accumulation in certain orientations in the two optical modes. Through experiments and theory, we show that the GAWBS-induced phase can be compensated down to the quantum-noise limit by self-referencing in a symmetric interferometer with Stokes detection.

Optical fiber-based interferometry has been an important tool in a variety of sensing applications. For example, the first ever demonstration of optical coherence tomography by Huang *et al.* [1] involved a fiber optic Michelson interferometer. Similarly, fiber optic gyroscopes [2], current sensors [3], hydrophone and magnetic sensors [4] have all incorporated Sagnac, Michelson, or Mach Zehnder interferometry depending upon the respective application. The phase sensitivity of fiber interferometers has been extensively maneuvered for handling non-classical light sources [5]. In particular, a symmetric nonlinear interferometer using Kerr squeezing can enable extremely sensitive noise reduction [5], [6]. The fiber ring reflector methodology, typically using a polarization-maintaining (PM) fiber in such a scenario, elegantly filters the squeezed signal and the pump into two separate arms for subsequent homodyne detection. However, in all forms of fiber interferometric systems, forward- or guided-acoustic-wave Brillouin scattering (GAWBS) from transverse acoustic phonons in the fiber disrupts the guided optical mode via phase modulation [7]. For instance, GAWBS degrades the performance of dense coding [8], entanglement swapping [9], and other essential components of quantum communication, and also limits high-speed classical communication [10], metrology, and interferometry [11] applications that require stable frequency, phase, or polarization control. In the case of nonclassical light, such as quadrature or photon-number squeezing, GAWBS sets the lower limit for the observation of squeezing [12], [13]. Essentially, at the advent of growing interest in quantum communication and networks, or delivering quantum sources between two points, thermal noise from effects such as GAWBS remains a major caveat.

There have been multiple attempts to reduce or counteract GAWBS in a fiber. For example, a natural approach would be to cool a fiber to the limit of phonon noise, as done by Shelby *et al.* [14]. However, scaling of systems involving liquid He at 4.2 K is inherently impractical. In a significant development, Elser *et al.* [15] showed that using a photonic crystal fiber (PCF), GAWBS can be substantially reduced up to hundreds of MHz. The microstructured air-hole photonic crystal in a PCF also enables tailoring the fiber parameters to decouple the optical and acoustic fields, reducing their overlap. But transverse phonon modes cannot be inherently removed as they spawn from the impedance mismatch between glass and the medium surrounding it. As a result, many practical applications rely on self-referencing of the light field. This can be intrinsically achieved in the symmetric interferometers used in quadrature squeezing and has hence been used to reduce GAWBS [16]. This is also reliable for achieving polarization-squeezing in a fiber using Kerr nonlinearity [17]. All these efforts with added experimental steps and challenges have mitigated GAWBS within the fiber platform, but never without a finite residual leftover. In this letter, we demonstrate for the first time the complete removal of GAWBS in continuous-wave laser operation in silica fiber using a twisted PCF (t-PCF). Twisted microstructured fibers offer robust symmetric eigenstates in the circular basis, resulting in azimuthal invariance to the polarization of light [18]. In the next section, we will develop the theoretical framework exploiting the symmetries offered by t-PCF and experimentally demonstrate its ability to completely remove GAWBS down to the quantum-noise limit. This novel feature, in addition to the endlessly single-mode guidance, tunable group velocity dispersion and nonlinearity, further adds to the prospect of the microstructured fiber platform.

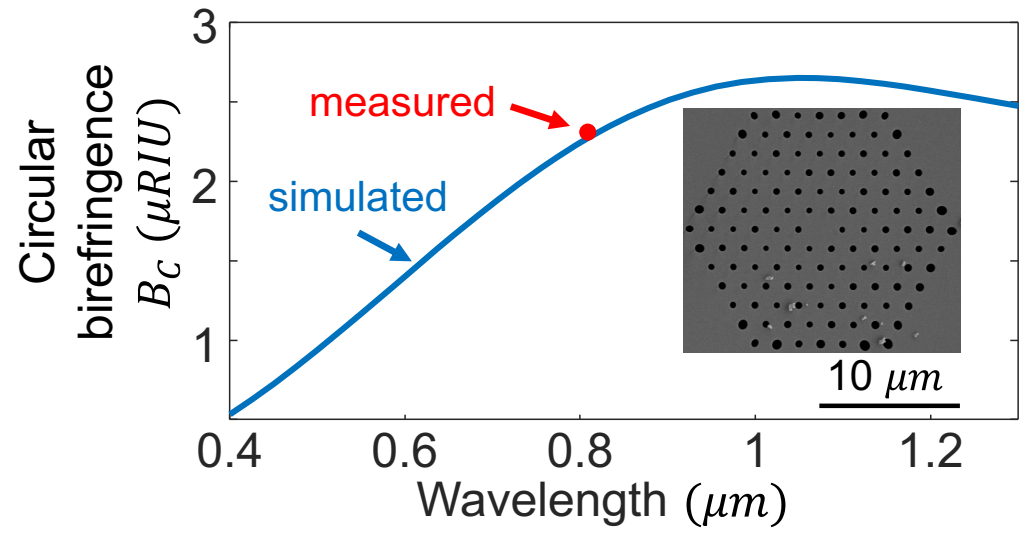

Fig. 1: The simulated circular birefringence ($B_C$) for the t-PCF in the SEM image (inset) in the laboratory frame is presented. The $B_C$ can be conveniently extracted from the topological Zeeman splitting as shown in Ref [19] in the twisted or laboratory frame. The t-PCF has a twist period of 3 mm, and the experimentally measured $B_C$ matches well with the value simulated at 810 nm using FEA.

Twisted PCF is fabricated either by drawing a PCF from a spinning preform or through fiber post-processing [18]. The helical trajectory of the air-holes along the length induces circular birefringence in the fiber through topological Zeeman splitting [19] in the twisted frame. The modal bifurcation results in circularly polarized eigenmodes $|R\rangle$ and $|L\rangle$ in the fiber. In Fig. 1, the simulated circular birefringence, $B_C$ at a twist period of 3 mm for the fiber in the scanning electron microscopy (SEM) image is presented. The finite element analysis (FEA) simulation predicts the $B_C$ accurately at 810 nm, as measured using the cutback method. In the Poincaré sphere representation, the fiber eigenstates are in the circular basis. Even in the weak birefringence limit (μRIU), the eigenmodes possess slightly different phase velocities from each other. When the in-coupling light is not purely circularly polarized, it is coupled into both eigenmodes simultaneously. In this case, the light field in the fiber can be parametrized as,

$$\tilde{E} = \left(cos\frac{\theta}{2}e^{-i\frac{\varphi}{2}}|R\rangle + sin\frac{\theta}{2}e^{i\frac{\varphi}{2}}|L\rangle\right) \qquad (1)$$

where $\theta$ and $\varphi$ are the polar and azimuthal angles respectively, in the Poincaré sphere in Fig. 2. During propagation, the light acquires a length-dependent phase, resulting in an azimuthal rotation of its polarization. The t-PCFs essentially preserve the $S_3$ Stokes parameter of any light field propagating in it, or the initial $\theta$ value remains invariant for any $\tilde{E}$. This property can be extremely useful, for instance, in quantum key distribution (QKD) applications [20], where the output polarization is confined to a one-dimensional protected subspace, simplifying many relevant utilities. Similarly, these symmetries offered by t-PCFs also result in low-noise circularly polarized supercontinuum [21], exotic polarization modulation instability [22], and vortex Brillouin laser [23]. Additionally, the $C_6$ rotational symmetry also offers rotational invariance of the in-coupled light polarization. This has a substantial advantage over commercially available PM fibers, which are extremely sensitive to the incoming polarization of light. The polarization noise at the quantum limit showed remarkable stability in t-PCFs over commercial PM-panda fibers [24].

In optical fibers, two types of GAWBS are commonly observed. The radial ($R_{0m}$) type modulates the fiber cross-section symmetrically, resulting in phase modulation of the guided mode. On the other hand, the mixed torsional radial ($TR_{2m}$) type modulates both phase and polarization. The phase and polarization modulation of the light field originates from the microscopic transverse vibrations of the optical fiber. This alters the refractive index ($n$) through the photoelastic effect described as

$$\Delta\left(\frac{1}{n^2}\right)_{ij} = p_{ijkl}S_{kl} \qquad (2)$$

where $p_{ijkl}$ are the coefficients of the photoelastic strain tensor and $S_{kl}$ the strain. Due to multiple symmetries and isotropy in silica, and a lack of z-component in GAWBS, there are only 2 independent parameters $p_{11}$, $p_{12}$ ($p_{44} = \frac{p_{11}-p_{12}}{2}$) present in it. As a result, the strain-induced change of refractive index $\Delta n(r, \emptyset)$ in the fiber's cylindrical frame is given as,

$$-\frac{n^3}{2}\begin{bmatrix} p_{11}\,S_{rr} + p_{12}\,S_{\emptyset\emptyset} & 2\,p_{44}\,S_{r\emptyset} \\ 2\,p_{44}\,\,S_{r\emptyset} & p_{12}\,S_{rr} + p_{11}\,S_{\emptyset\emptyset} \end{bmatrix}$$

In the circular basis, the $\Delta n(r, \emptyset)$ further becomes

$$\Delta n_{R/L} = \begin{bmatrix} N & g \\ g^* & N \end{bmatrix} \qquad (3)$$

where $N = -\frac{n^3}{2}(p_{11}p_{12})(S_{rr} + S_{\emptyset\emptyset})$ and $g = -n^3(p_{44}(S_{rr} - S_{\emptyset\emptyset}) + ip_{44}\,S_{r\emptyset}$. Here the diagonal $N$ terms contribute to pure phase modulation, and their

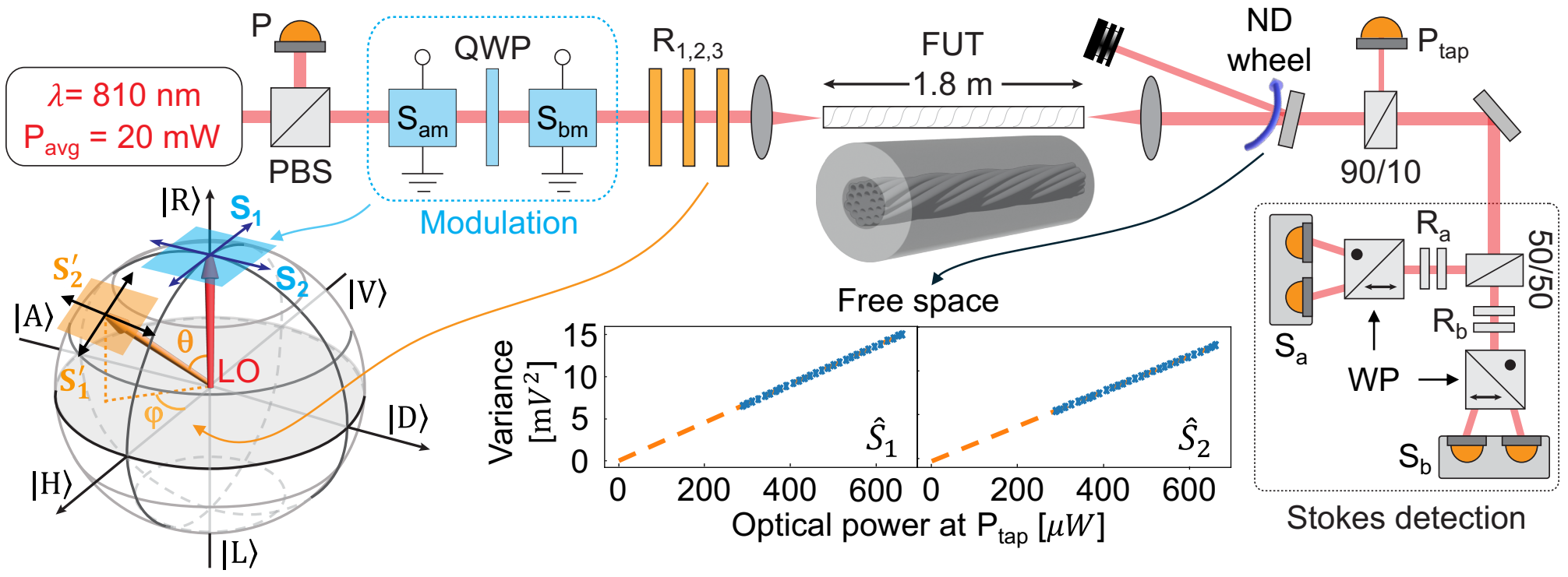

Fig. 2: The experimental setup employs a low-noise laser at 810 nm for signal and LO preparation. A pair of polarization modulators $S_{am}$ and $S_{bm}$ are used to align the Stokes detectors respectively, allowing the measurement of polarization noise in the dark plane orthogonal to the LO. The detectors are low-noise, home-built, balanced detectors with a frequency response of ~70 MHz (3 dB). Using the retarders $R_1$, $R_2$, and $R_3$, the polarization states of the LO and the measurement plane can be chosen at different positions on the Poincaré sphere and rotated arbitrarily within it. The two stages are linked through a 1.8-meter t-PCF with 10 mW power (~50 % coupling efficiency) coupled into it. The fiber does not need to be kept straight because of the robustness of its circular birefringence. Before measuring the fiber's properties, the vacuum variance of free space was calibrated through attenuation measurement using the ND wheel. The linearity of the variances of $\hat{S}_1$ and $\hat{S}_2$ validates the shot noise-limited operation of the setup. (Abbreviations used: PBS: polarizing beam splitter; QWP: quarter-wave plate; ND filter: neutral density filter; FUT: fiber under test; R: Retarder; P: power monitor; WP: Wollaston prism)

complete analytical expressions in a silica fiber were calculated by Shelby *et al.* [25]

To observe the $TR_{2m}$ GAWBS modes in a t-PCF, we use the Stokes detection setup shown in Fig. 2. A narrow-linewidth (<1 MHz) DLpro series laser from Toptica at 810 nm provides the transmitted bright local oscillator (LO). $S_{am}$ and $S_{bm}$, together with a QWP, are used to modulate signal states onto the LO. For a bright LO and small amplitudes, the resulting signal states can be visualized on a tangential plane on the Poincaré sphere, orthogonal to the LO. In our system, the QWP is chosen so that the modulation of $S_{am}$ and $S_{bm}$ define the orthogonal axes of this tangential plane. The modulated signal states are used for aligning the Stokes detection units $S_a$ and $S_b$ before a measurement to the axes of the tangential plane. The LO and the resulting measurement plane are depicted with the red arrow pointed towards the north pole and the blue plane atop it in the Poincaré sphere in Fig. 2, respectively. The LO and signal states exist in the same spatial mode but in orthogonal polarization modes. The quantum mechanical description of the input state in the circular basis can be given as [26],

$$|\psi\rangle = |\alpha_{LO}\rangle_R |\alpha_{sig}\rangle_L$$

Since the fiber's eigenstates are in the circular basis, the signal and LO are also defined in the R/L basis with $\hat{\alpha}_{L/R} = \frac{1}{\sqrt{2}}(\hat{\alpha}_x \pm i\hat{\alpha}_y)$. But in principle, they can be defined in any orthonormal basis. The subsequent quantum Stokes operators can then be defined as [27]

$$\hat{S}_0 = \hat{\alpha}_R^\dagger \hat{\alpha}_R + \hat{\alpha}_L^\dagger \hat{\alpha}_L \,; \quad \hat{S}_1 = \hat{\alpha}_R^\dagger \hat{\alpha}_L + \hat{\alpha}_L^\dagger \hat{\alpha}_R$$
$$\hat{S}_2 = i(\hat{\alpha}_L^\dagger \hat{\alpha}_R - \hat{\alpha}_R^\dagger \hat{\alpha}_L) \,; \quad \hat{S}_3 = \hat{\alpha}_R^\dagger \hat{\alpha}_R - \hat{\alpha}_L^\dagger \hat{\alpha}_L \quad (4)$$

Thus, for the orientation shown in blue in Fig. 2, the alignment procedure leads to a measurement of $\hat{S}_1$ and $\hat{S}_2$. After the alignment, the modulators are turned off, making the signal mode in the vacuum state $|0\rangle$ in the orthogonal polarization mode to the LO [26]. With the help of $R_{1,2,3}$, the LO can be oriented along any point in the Poincaré sphere to probe arbitrary $\theta$ and $\varphi$ values. Thus, the resulting Stokes detection allows probing of the polarization noise in the vacuum state at any point on the Poincaré sphere. For $\theta \neq 0$ and $\varphi \neq 0$, the corresponding measurements are indicated along $S_1'$ and $S_2'$ on the local measurement plane (depicted in yellow in Fig. 2).

Stokes detectors $S_a$ and $S_b$ essentially measure the interference of vacuum and LO modes at the respective WP. When the measurement is conducted for light propagation through air (fiber is removed), the signal and LO do not acquire any relative phase as air is not birefringent. This leads to the case of symmetric linear interferometry. Additionally, a power-dependent variance measurement of the vacuum mode without the fiber shows linear dependence of the variances of $S_a$ and $S_b$. This is an affirmation of the absence of any classical noise in the system as a whole.

To measure the polarization noise in a fiber, the t-PCF in Fig. 1 with $B_C$ of 2.3 μRIU and outer diameter of 250 μm is mounted on the setup. Within the 70 MHz bandwidth of the detectors, 6 $TR_{2m}$ GAWBS modes are observed in the fiber as expected (Fig. 3). Since $|0\rangle$ and $|\alpha_{LO}\rangle$ are always in the same spatial mode, phase modulation arising from $R_{0m}$ modes get auto-compensated at detection and hence are not observed. Thus, for all the measurements concerned using the setup, we will treat $N = 0$ in Eq. (3).

In Fig. 3, three different power spectrum measurements of the Stokes detectors are shown with the respective $\theta$ and $\varphi$ angles in the Poincaré sphere depiction. In all cases, the noise floor for the free-space reference measurement indicates the shot noise of the vacuum mode orthogonal to LO. In the fibers, the GAWBS peaks are observed when $\theta = 0^0$. However, when $\theta = 90^0$, the peaks disappear at the detector $S_b$. In these configurations, the noise floor completely aligns with free space, indicating complete absence of any thermal noise. None of the earlier reports mention the disappearance of GAWBS in either linear or nonlinear interferometry.

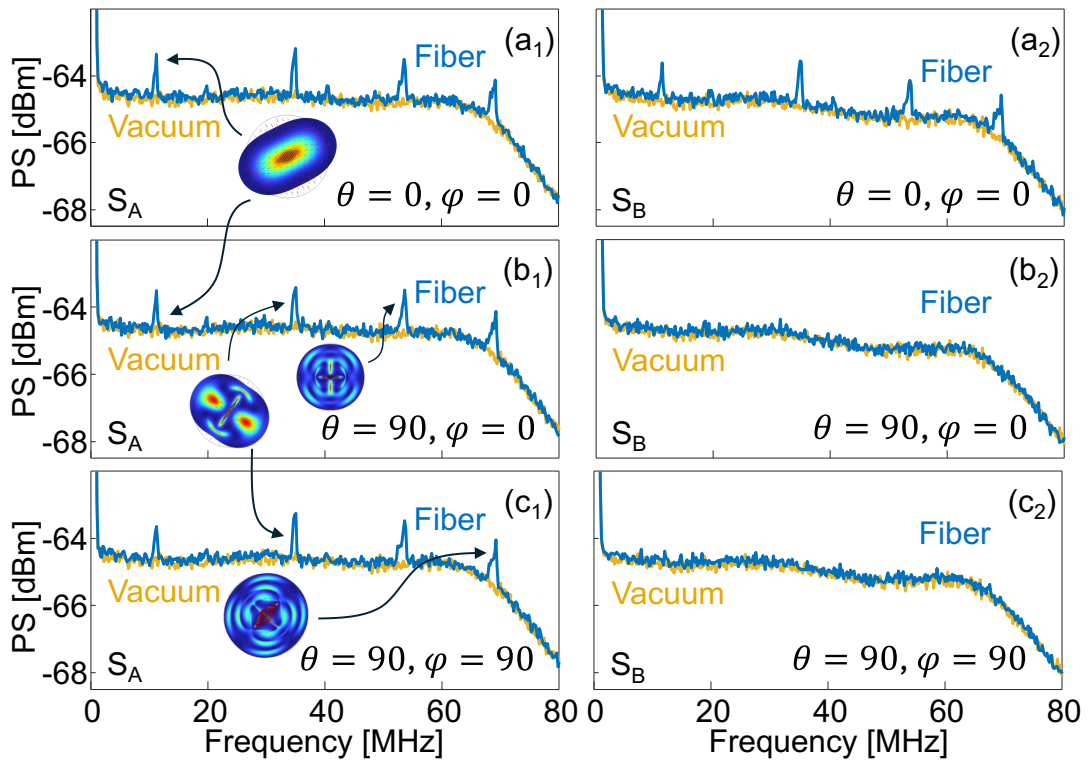

Fig. 3: (a), (b) and (c) show the power spectrum at the $S_a$ and $S_b$ detectors for 3 different positions of the LO along $|R\rangle$, $|H\rangle$, $|D\rangle$ respectively. In (a1) and (a2), the $TR_{2m}$ GAWBS modes appear atop the vacuum reference. In (b2) and (c2), the GAWBS peaks vanish at $S_b$. Representative strain energy density for the 4 dominant GAWBS modes simulated using FEA is shown (2 additional modes at 19.6 MHz and 41 MHz are very close to the noise floor).

The GAWBS disappearance phenomena in t-PCFs can be understood as follows. Since GAWBS is a classical effect and measured atop the quantum noise, we will treat the theoretical formalism semi-classically. A light field $E = E_0 \exp(ikz)$ after propagating through a t-PCF of length $l$ emerges as $E_0 \exp(ikl\Delta n_{R/L})$ due to the phase acquisition from GAWBS. Using Eq. (1), we can use the Jones matrix formalism to express the field before ($E_{in}$) and after ($E_{out}$) the fiber in the circular basis as

$$E_{in} = \begin{bmatrix} cos\frac{\theta}{2}e^{-i\frac{\varphi}{2}} \\ sin\frac{\theta}{2}e^{i\frac{\varphi}{2}} \end{bmatrix} E_0$$

$$E_{out} = \left[\begin{bmatrix} 1 & 0 \\ 0 & 1 \end{bmatrix} + i\begin{bmatrix} 0 & g \\ g^* & 0 \end{bmatrix}\right] \begin{bmatrix} cos\frac{\theta}{2}e^{-i\frac{\varphi}{2}} \\ sin\frac{\theta}{2}e^{i\frac{\varphi}{2}} \end{bmatrix} E_0$$

Here we are ignoring the higher-order perturbations in the refractive index modulation, i.e. $\Delta n^2 \approx 0$ and $|g|^2 \approx 0$. The identity matrix appears due to the aligning of $S_a$ and $S_b$ detectors along the local measurement plane before each measurement. This balances the global phase acquired by the light due to the $B_C$ in the fiber, simplifying the theoretical model significantly. Similarly, since the laser's wavelength and fiber length are constant, we absorb them into $\Delta n_{R/L}$. Thus, the final expression for $E_{out}$ is

$$\begin{bmatrix} E_R \\ E_L \end{bmatrix} = \begin{bmatrix} E_0\left(cos\frac{\theta}{2}e^{-i\frac{\varphi}{2}} + ig\, sin\frac{\theta}{2}e^{i\frac{\varphi}{2}}\right) \\ E_0\left(sin\frac{\theta}{2}e^{i\frac{\varphi}{2}} + ig^*\, cos\frac{\theta}{2}e^{-i\frac{\varphi}{2}}\right) \end{bmatrix}$$

Assuming $|E_0|^2 = P$ we can evaluate the following

$$E_R^* E_L = P\left(\frac{sin\,\theta}{2}e^{i\frac{\varphi}{2}} + ig^*\, cos\,\theta\right)$$
$$|E_R|^2 = P\left(cos^2\frac{\theta}{2} + \frac{i}{2}sin\,\theta\,(ge^{i\varphi} - g^*e^{-i\varphi})\right)$$
$$|E_L|^2 = P\left(sin^2\frac{\theta}{2} - \frac{i}{2}sin\,\theta\,(ge^{i\varphi} - g^*e^{-i\varphi})\right)$$

Using the real and imaginary parts of $g = g' + ig''$, we can evaluate the Stokes parameters of the $E_{out}$ field using the classical analogue of Eq. (4). Thus,

$$S_1 = 2Re(E_R^* E_L) = P\left(sin\,\theta\, cos\frac{\varphi}{2} + 2g''\, cos\,\theta\right)$$
$$S_2 = 2Im(E_R^* E_L) = P\left(sin\,\theta\, sin\frac{\varphi}{2} + 2g'\, cos\,\theta\right)$$
$$S_3 = |E_R|^2 - |E_L|^2 = P\big(cos\,\theta - 2\,sin\,\theta\,(g'\,sin\,\varphi + g''\,cos\,\varphi)\big)$$
$$S_0 = |E_R|^2 + |E_L|^2 = P \qquad (5)$$

Finally, these expressions for the Stokes parameters are defined for the measurement plane at the north pole of the Poincaré sphere (blue plane on Fig. 2 corresponding to $\theta = 0$). For any arbitrary orientation of the LO ($\theta \neq 0$) along the first meridian ($\varphi = 0$), the measurement plane can be conveniently written using a rotation matrix for $\theta$

$$\begin{bmatrix} S_1' \\ S_2' \\ S_3' \end{bmatrix} = \begin{bmatrix} \cos\theta & 0 & -sin\,\theta \\ 0 & 1 & 0 \\ sin\,\theta & 0 & cos\,\theta \end{bmatrix} \begin{bmatrix} S_1 \\ S_2 \\ S_3 \end{bmatrix}$$

to obtain,

$$\begin{bmatrix} S_1' \\ S_2' \\ S_3' \end{bmatrix} = \begin{bmatrix} 2g''P \\ -2Pg'\cos\theta \\ P \end{bmatrix}$$

and the total power is unchanged, so that $S_0' = S_0$. The subsequent variances of the new Stokes parameters are,

$$\begin{bmatrix} var(S_1') \\ var(S_2') \\ var(S_3') \end{bmatrix} = \begin{bmatrix} 4P^2\Delta g'' \\ 4P^2\Delta g' \cos^2\theta \\ 0 \end{bmatrix} \qquad (6)$$

where $\Delta g'$ and $\Delta g''$ are the variances in the real and imaginary parts of $g$ respectively. And the variance of $P$ is null, $var(S_0') = 0$.

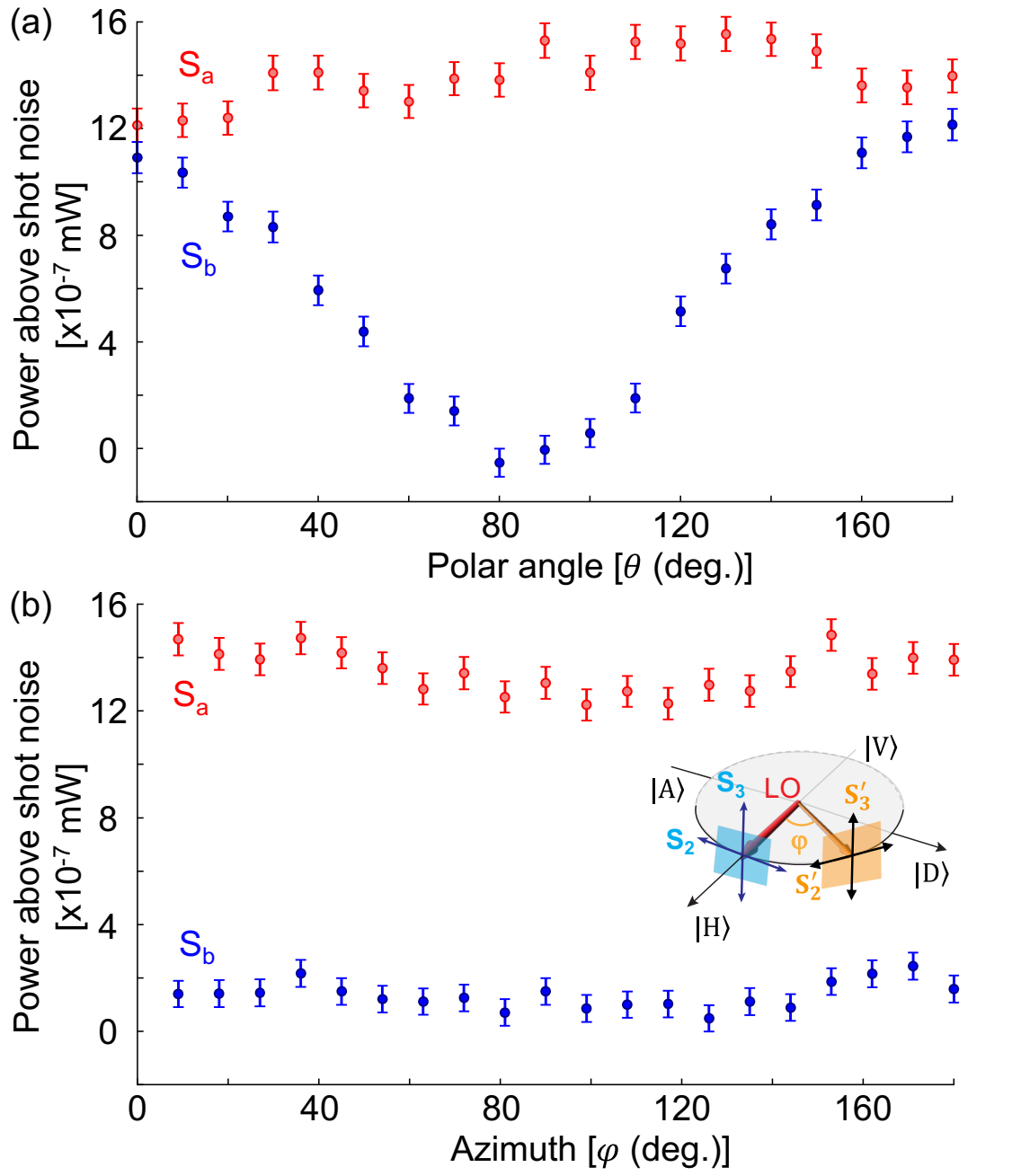

Fig. 4: The variation of total GAWBS power for moving the LO along the first meridian of the Poincaré sphere is shown in (a). As predicted from Eq. (6), the GAWBS power scales as the square of the cosine in $S_b$ while maintaining a fixed value at $S_a$. When the measurement is swept along the equator of the Poincaré sphere, the GAWBS powers remain unchanged at $S_a$ and $S_b$ as shown in (b).

Fig. 4 (a) shows the experimentally measured total GAWBS power as the LO is varied along the first meridian. We integrated the measured power spectra over 1 MHz windows around the 4 dominant GAWBS resonances and subtracted the corresponding vacuum reference. The integrated power is proportional to the signal's variance in the same bandpass-filter frequency windows. $S_a$ and $S_b$ measure $var(S_1')$ and $var(S_2')$ respectively. As the polar angle $\theta$ progresses from $0^0$ to $180^0$, the GAWBS remains unchanged at the $S_a$ detector. This replicates the lack of $\theta$ dependence in $var(S_1')$. However, $S_b$ follows the sinusoidal evolution along the meridian, exhibiting zero GAWBS noise at the equator ($\theta = 90^0$). Additionally, at $\theta = 0^0, 180^0$, the experimentally measured GAWBS at $S_a$ and $S_b$ are observed to be equal. In t-PCFs, the structural symmetry evidently compels $g'$ and $g''$ to be the same. $g'$ and $g''$ have a phase difference of $90^0$ and they represent the doubly degenerate GAWBS modes which are identical when the structure has azimuthal symmetry. This can be easily simulated using FEA tools for PCFs, t-PCFs, or even symmetric step-index fibers. This can also be inferred from the doubly degenerate nature and analytical expressions derived by Shelby *et al.* [25]. Additionally, the experimentally observed similar variances of $S_a$ and $S_b$ at $\theta = 0^0, 180^0$ also indicates $\Delta g'$ and $\Delta g''$ to be equal. At $\theta = 0^0$, Eq. (6) gives, $var(S_1') = var(S_1) = 4P^2\Delta g''$ and $var(S_2') = var(S_2') = 4P^2\Delta g'$. Further, if we keep $\theta$ fixed at 90° and rotate the LO along the equator of the Poincaré sphere, we obtain the traces in Fig. 4 (b). In this configuration, the $S_a$ and $S_b$ detectors measure $var(S_3')$ and $var(S_2')$ respectively. Using Eq. (5) with an appropriate rotation matrix for $\varphi$, we can similarly derive,

$$var(S_2') = 0$$
$$var(S_3') = 4P^2(\Delta g' \sin^2\varphi + \Delta g'' \cos^2\varphi)$$

which is at par with the experimental result. However, this also directly implies that $\Delta g' = \Delta g''$. The geometric azimuthal invariance in the cross-section of t-PCFs can be applied to both optical as well as the acoustic modes, making $S_1$ and $S_2$ indistinguishable.

The interchangeability of $S_1$ and $S_2$ can also result in unique consequences. For example, for LO at $\theta = 90°, \varphi = 0$, the modulation plane itself can be rotated about the LO-axis using $R_{1,2,3}$ in Fig. 2. The orientation of the measurement plane redistributes the phase accumulation between $S_1$ and $S_2$ leading to the double helix trace from Fig. 5 (a). We define the orientation angle $\psi$ as the angle of $S_2'$ from the usual $S_2$ direction on the orthogonal plane in degrees (inset Fig. 5 (a)). Further, if we vary the LO along the first meridian ($\varphi = 0$) similar to Fig. 4 (a) (yellow arrows in Fig. 5(b) inset), but simultaneously also rotate $\psi$ from $0^0$ to $180^0$, both Stokes detectors exhibit GAWBS reduction simultaneously, although not indefinitely in $S_a$.

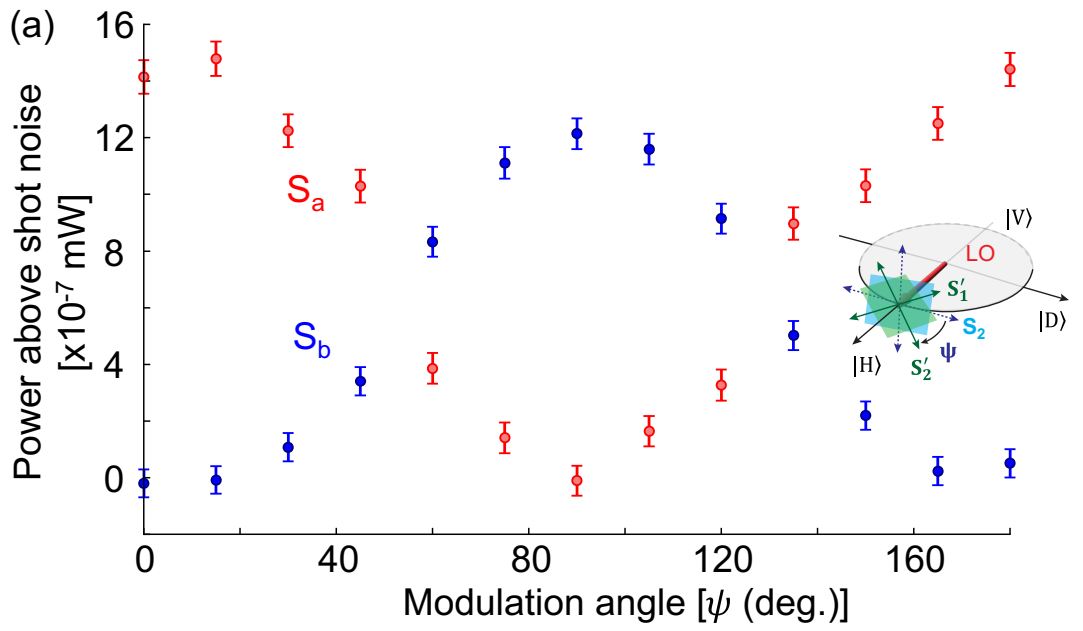

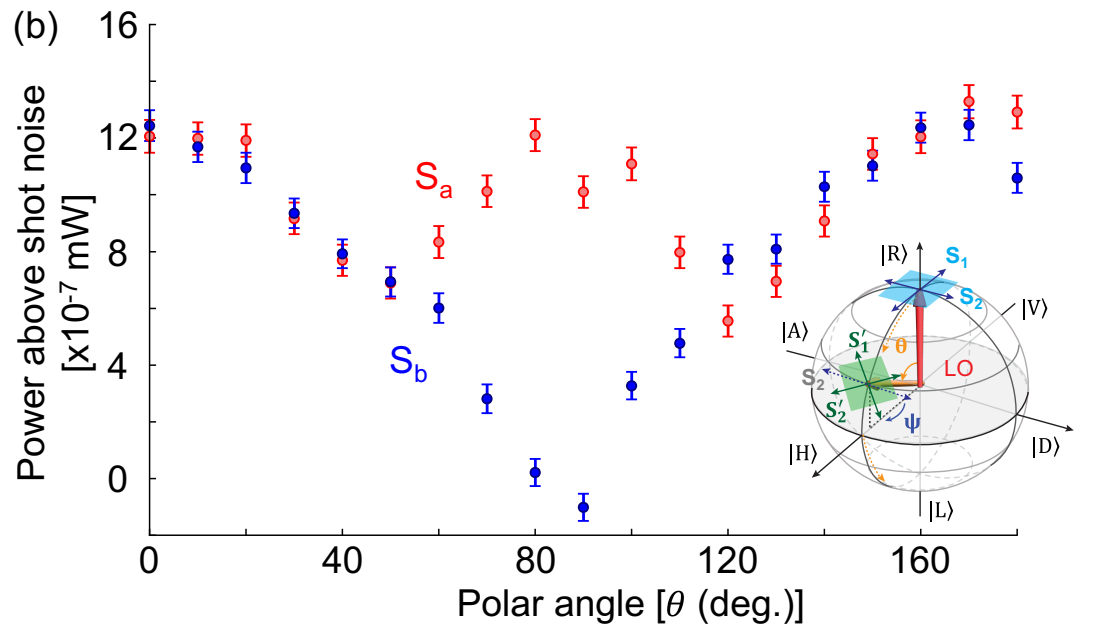

Fig. 5: The variation of total GAWBS power when the modulation plane is rotated, maintaining the LO at $|H\rangle$ is shown in (a). This shows the interchangeability of $S_1$ and $S_2$ in t-PCFs. In (b), the outcome for simultaneous sweeping of the LO from $|R\rangle$ to $|L\rangle$ while rotating the modulation plane is presented.

These features offered by t-PCFs are highly promising for their prospects in fiber-based interferometry. For quadrature or polarization squeezing, we expect that all previously implemented methodologies will be highly simplified. These features are not possible to achieve in other PM fiber systems. Especially in the linearly birefringent systems, where the structure has a strong broken azimuthal symmetry. This will lead to uneven strain energy in the phonon modes leading to unequal $g'$ and $g''$ and possibly unequal $\Delta g'$ and $\Delta g''$ as well. t-PCFs can offer channels devoid of classical noise and with suitable protocols they can be used to interface between CV-QKD modules in short-range transmissions.

In summary, we have demonstrated GAWBS mitigation down to the quantum noise limit for the first time in an optical fiber. This is achieved by exploiting the robust symmetric optical eigenmodes and their equal overlap with the acoustic phonons in a twisted PCF. The equality of the overlap originates from the geometric symmetry in the fiber cross section. In contrast, these symmetries cannot exist in a linearly birefringent PM fiber, as the reduced cross-talk of the optical modes requires strong broken symmetry in the geometry. This presents t-PCFs as a highly potent fiber platform for various applications in quantum optics. Additionally, for short-range point-to-point connections in free-space CV-QKD infrastructure that employs polarization bases, t-PCFs can provide silent noiseless channels. While the cost and procedure make them less likely for scalable deployment in long-range telecommunication systems, they can be useful in short-range operations such as in data centers. Twisted PCFs will be highly desirable for shot-noise-limited fiber interferometric applications. Especially for quantum-enhanced sensing using quadrature or polarization-squeezed light, a classical-noise-free fiber will provide both generation and delivery simultaneously.

[1] D. Huang, E. A. Swanson, C. P. Lin, J. S. Schuman, W. G. Stinson, W. Chang, M. R. Hee, T. Flotte, K. Gregory, C. A. Puliafito, and J. G. Fujimoto, Optical Coherence Tomography, Science **254**, 5035 (1991).

[2] H. J. Arditty and H. C. Lefèvre, Sagnac effect in fiber gyroscopes, Opt. Lett. **6**, 8 (1981).

[3] G. Frosio and R. Dändliker, Reciprocal reflection interferometer for a fiber-optic Faraday current sensor, Appl. Opt. **33**, 25 (1994).

[4] T. G. Giallorenzi, J. A. Bucaro, A. Dandridge, G. H. Sigel, J. H. Cole, S. C. Rashleigh, and R. G. Priest, Optical Fiber Sensor Technology, IEEE Trans. on Microwave Theory and Techniques, **30**, 4 (1982).

[5] M. Shirasaki, and H. A. Haus, Squeezing of pulses in a nonlinear interferometer, J. Opt. Soc. Am. B **7**, 1 (1990).

[6] K. Bergman, and H. A. Haus, Squeezing in fibers with optical pulses, Opt. Lett. **16**, 9 (1991).

[7] R. M. Shelby, M. D. Levenson, and P. W. Bayer, Resolved forward Brillouin scattering in optical fibers, Phys. Rev. Lett. **54**, 939 (1985).

[8] S. L. Braunstein and H. J. Kimble, Dense coding for continuous variables, Phys. Rev. A **61,** 042302 (2000).

[9] O. Glöckl, S. Lorenz, Ch. Marquardt, J. Heersink, M. Brownnutt, C. Silberhorn, Q. Pan, P. van Loock, N. Korolkova, and G. Leuchs, Experiment towards continuous-variable entanglement swapping: Highly correlated four-partite quantum state, Phys. Rev. A **68**, 012319 (2003).

[10] M. Nakazawa, M. Yoshida, M. Terayama, S. Okamoto, K. Kasai, and T. Hirooka, Observation of guided acoustic-wave Brillouin scattering noise and its compensation in digital coherent optical fiber transmission, Opt. Exp. **26**, 7 (2018).

[11] C. Hilweg, D. Shadmany, P. Walther, N. Mavalvala, and V. Sudhir, Limits and prospects for long-baseline optical fiber interferometry, Optica **9**, 11 (2022).

[12] K. Bergman, H. A. Haus, and M. Shirasaki, Analysis and measurement of GAWBS spectrum in a nonlinear fiber ring, App. Phys. B. **55**, 242-249 (1992).

[13] A. Hosaka, K. Hirosawa, R. Sawada, and F. Kannari, Generation of photon-number squeezed states with a fiber-optic symmetric interferometer, Opt. Exp. **23**, 15 (2015).

[14] R. M. Shelby, M. D. Levenson, S. H. Perlmutter, R. G. DeVoe, and D. F. Wallst'l, Broad-Sand Parametric Deamplification of Quantum Noise in an Optical Fiber, Phys. Rev. Lett. **57**, 6 (1986).

[15] D. Elser, U. L. Andersen, A. Korn, O. Glöckl, S. Lorenz, Ch. Marquardt, and G. Leuchs, Reduction of Guided AcousticWave Brillouin Scattering in Photonic Crystal Fibers, Phys. Rev. Lett., **97**, 133901 (2006).

[16] N. Kalinin, T. Dirmeier, A. A. Sorokin, E. A. Anashkina, L. L. Sánchez-Soto, J. F. Corney, G. Leuchs, and A. V. Andrianov, Quantum-enhanced interferometer using Kerr squeezing Nanophot. **12**, 14 (2023).

[17] N. Kalinin, T. Dirmeier, A. A. Sorokin, E. A. Anashkina, L. L. Sánchez-Soto, J. F. Corney, G. Leuchs, A. V. Andrianov, Observation of robust polarization squeezing via the Kerr nonlinearity in an optical fiber, Adv. Quantum Technol. **6**, 3 (2023).

[18] P. St. J. Russell, R. Beravat, and G. K. L.Wong, Helically twisted photonic crystal fibres, Phil. Trans. R. Soc. A **375**, 2087 (2017).

[19] T. Weiss, G. K. L. Wong, F. Biancalana, S. M. Barnett, X. M. Xi, and P. St. J. Russell, Topological Zeeman effect and circular birefringence in twisted photonic crystal fibers, J. Opt. Soc. Am. B **30**, 11 (2013).

[20] M. A. T. Butt, P. Roth, G. K. L. Wong, M. H. Frosz, L. L. Sánchez-Soto, E. A. Anashkina, A. V. Andrianov, P. Banzer, P. St. J. Russell, and G. Leuchs, Protecting Quantum Modes in Optical Fibers, Phys. Rev. App. **19**, 054080 (2023).

[21] M. Lippl, M. H. Frosz, and N. Y. Joly, Low-noise supercontinuum generation in chiral all-normal dispersion photonic crystal fibers, Opt. Lett. **48**, 20 (2023).
[22] P. Roth, P. S. J. Russell, M. H. Frosz, Y. Chen and G. K. L. Wong, Modulational Instability and Spectral Broadening of Vortex Modes in Chiral Photonic Crystal Fibers, Journ. of Lightwave Technol. **41**, 7 (2023).
[23] X. Zeng, P. St. J. Russell, Y. Chen, Z. Wang, G. K. L. Wong, P. Roth, M. H. Frosz, and B. Stiller, Optical Vortex Brillouin Laser, Laser Photonics Rev. **17**, 4 (2023).
[24] V. Choudhury, K. Jaksch, M. Lippl, M. Frosz, W. Elser, G. Leuchs, Ch. Marquardt, and N. Joly, Quantum-Limited Polarization Measurements in Twisted Photonic Crystal Fibers, in CLEO 2025, Technical Digest Series (Optica Publishing Group, 2025), paper FF120_2.
[25] R. M. Shelby, M. D. Levenson, and P. W. Bayer, Guided acoustic-wave Brillouin scattering, Phys. Rev. B **31**, 8 (1985).
[26] K. Jaksch, T. Dirmeier, Y. Weiser, S. Richter, Ö. Bayraktar, B. Hacker, C. Rösler, I. Khan, S. Petscharning, T. Grafenauer, M. Hentschel, B. Ömer, Ch. Pacher, F. Kanitschar, T. Upadhyaya, J. Lin, N. Lütkenhaus, G. Leuchs, and Ch. Marquardt, Composable free-space continuous-variable quantum key distribution using discrete modulation, Sci. Adv. **12**, 24 (2026).
[27] N. Korolkova, G. Leuchs, R. Loudon, T. C. Ralph, and C. Silberhorn, Polarization squeezing and continuous-variable polarization entanglement, Phys. Rev. A **65**, 052306 (2002).